\pdfoutput=1
\begingroup\expandafter\expandafter\expandafter\endgroup
\expandafter\ifx\csname pdfsuppresswarningpagegroup\endcsname\relax
\else
\fi

\documentclass[final]{elsarticle} 

\usepackage[utf8]{inputenc}

\usepackage[hyphens]{url}
\biboptions{sort&compress, square, comma}
\bibpunct{{\unskip~[}}{]}{,}{n}{}{;} 
\usepackage{threeparttable}

\usepackage[breaklinks=true, linkcolor=blue, citecolor=blue, colorlinks=true]{hyperref}
\usepackage{graphicx}
\usepackage{caption,subcaption}
\usepackage[version=4]{mhchem}
\usepackage{booktabs,multirow}
\usepackage{latexsym,amsmath,amssymb}
\usepackage[section]{placeins} 
\usepackage[textsize=small]{todonotes}  

\usepackage[T1]{fontenc}
\usepackage[english]{babel}
\usepackage{csquotes}
\usepackage{textcomp}

\usepackage[binary-units]{siunitx}
\DeclareSIUnit\atm{atm}
\DeclareSIUnit\bar{bar}
\DeclareSIUnit{\calorie}{cal}

\usepackage{array}
\journal{Combustion and Flame}

\begin{document}

\begin{frontmatter}

\title{Assessing impacts of discrepancies in model parameters\\ on autoignition model performance:\\ a case study using butanol}

\author[neu]{Sai~Krishna~Sirumalla}
\author[osu]{Morgan~A.~Mayer}
\author[osu]{Kyle~E.~Niemeyer}
\author[neu]{Richard~H.~West\corref{cor1}}
\ead{r.west@northeastern.edu}

\address[neu]{Department of Chemical Engineering\\
Northeastern University, Boston, MA 02115, USA}
\address[osu]{School of Mechanical, Industrial, and Manufacturing Engineering\\
Oregon State University, Corvallis, OR 97331, USA}

\cortext[cor1]{Corresponding author}

\begin{abstract}
Side-by-side comparison of detailed kinetic models using a new tool to aid recognition of species structures reveals significant discrepancies in the published rates of many reactions and thermochemistry of many species.
We present a first automated assessment of the impact of these varying parameters on observable quantities of interest---in this case, autoignition delay---using literature experimental data.
A recent kinetic model for the isomers of butanol was imported into a common database.
Individual reaction rate and thermodynamic parameters of species were varied using values encountered in combustion models from recent literature.
The effects of over 1,600 alternative parameters were considered.
Separately, experimental data were collected
from recent publications
and converted into the standard YAML-based ChemKED format.
The Cantera-based model validation tool, PyTeCK, was used to automatically simulate autoignition using the generated models and experimental data, to judge the performance of the models.
Taken individually, most of the parameter substitutions have little effect on the overall model performance, although a handful have quite large effects, and are investigated more thoroughly.
Additionally, models varying multiple parameters simultaneously were evolved using a genetic algorithm to give fastest and slowest autoignition delay times, showing that changes exceeding a factor of 10 in ignition delay time are possible by cherry-picking from only accepted, published parameters.
All data and software used in this study are available openly.
\end{abstract}


\end{frontmatter}


\section{Introduction}
Detailed kinetic models over a range of temperatures and pressures are essential for predicting the behavior of new fuels.
Kinetic combustion models of complicated fuels contain thousands of species and elementary reactions which are described by thermodynamic and rate parameters.
Many of these parameters are calculated with semi-empirical methods, estimated,  sometimes just guessed, and quite often changed or ``tweaked'' to alter some global observable.
This leads to discrepancies in rates and thermodynamic parameters for the same reaction or species in different models.
The work presented aims to determine how these discrepancies affect the performance of a model.

Side-by-side comparison of detailed kinetic models reveals significant discrepancies in the published rates of many reactions and thermochemistry of many species.
For example, in the supplementary data of the 2016 Combustion Symposium proceedings, of 2,600 reactions we have identified in two or more models, 15\% disagree by over an order of magnitude at 1000~K, and some by 31 orders of magnitude;
of the species we found in two or more models, 4\% of standard enthalpy of formation values span more than \SI{50}{\kilo\joule\per\mole}.
{Chen et al.~[\citenum{CHEN2017208}] recently used an automated tool to show that many published models have rate coefficients exceeding the collision limit by several orders magnitude.}
However, the impact of these variations on observable quantities of interest---such as autoignition delay---has not yet been assessed.
Each published model has usually been ``validated'' with and often trained, optimized, or tweaked to match a given set of experimental data.
Many reaction rates have been chosen only as part of a whole model and only to match a limited set of experimental data, although they are then frequently used in other models.

Pioneering work by Frenklach and colleagues~\cite{Frenklach:1992}
advanced the systematic treatment of kinetic parameter uncertainty in combustion modeling.
Other notable contributions include those by
Wang~\cite{WangSheen2015},
Tur\'anyi~\cite{Turanyi:2014df}, and Tomlin~\cite{Tomlin2013}, whose reviews, books, and chapters provide a thorough and clear overview of local and global uncertainty analysis in this field.

Recent advances include treatment of correlations between uncertain parameters  derived from a common rate rule~\cite{Fridlyand:2017ki}
and the use of  multi-scale informatics~\cite{Burke:2016kd} to propagate uncertainties from physically meaningful molecular properties rather than reaction pre-exponential factors.
Many approaches involve Monte Carlo sampling within a range of uncertainties, attributed to every parameter by hand or according to some heuristics.
However, the systematic assessment of how much uncertainty could be due to discrepancies between parameters in published models has not been attempted, not because the mathematics are complicated but because the data are scattered and hard to reconcile into a common platform.
Because species are given different names in different models, it can be hard to find the discrepancies.

In this work we use butanol as a case study.
Bio-butanol is a potential renewable biofuel, offering several advantages over bio-ethanol:
its higher heating value allows a higher blending rate in gasoline;
its lower latent heat of vaporization reduces issues associated with combustion cold start~\cite{2011-01-0902};
it is less corrosive, has a higher cetane number, and lower vapor pressure; and
it has a similar viscosity to diesel.
Butanol research is still of interest to the combustion kinetics community, although not so novel as to be without data for comparison.
As well as a popular validated model from Lawrence Livermore National Laboratories by Sarathy et al.~\cite{Sarathy:2012fj}, upon which we base our investigation,
there are plenty of experimental data~\cite{Moss:2008bva,Stranic:2012jl,Bec:2014,Zhu:2014td}.
Tomlin et al.~\cite{AGBRO2017776} recently investigated the Sarathy et al.\ model used in this work by conducting both local and global uncertainty and sensitivity methods for predicting autoignition delay times and species profiles.

\section{Methods}

The overall workflow is to take {an original} model (the LLNL butanol model~\cite{Sarathy:2012fj}) in \textsc{Chemkin} format, and for the rate of every reaction rate and the thermochemistry of every species, search to see if an alternative has been used in any other recently published kinetic models. This gives a large set of alternative parameters, each of which has been independently ``validated,'' ``approved,'' or at least shared with the community.
In one analysis, we consider each variation independently, and measure its impact on the model performance as judged against a broad set of experimental data (475 datapoints across 67 datasets from four papers \cite{Moss:2008bva,Stranic:2012jl,Bec:2014,Zhu:2014td}).
This allows us to rank the parameters about which there exist disagreement, in order of importance for ignition delay predictions.
In a second analysis, we allow many parameters to be varied simultaneously using a genetic algorithm to explore the extrema---fastest and slowest ignition delays---that can be achieved by selecting from among the published alternative parameters.

\subsection{Kinetic model curation}
The major barrier to comparing published kinetic models is the lack of canonical or even conventional methods of naming the chemical species, combined with the persistent use of a \textsc{Chemkin}-compatible data structure designed not to preserve chemical metadata but rather to fit on an 80-column punch card. This has led to many alternative names being used for each species.
For example, {prop-1-en-2-yl} has been referred to in published models as \texttt{C*C.C}, \texttt{tC3H5}, \texttt{CH2CCH3}, \texttt{propen2yl}, \texttt{ch3cch2}, \texttt{T-C3H5}, \texttt{CH3CCH2}, \texttt{TC3H5}, \texttt{C3H5-T}, and \texttt{c3h5-t}, making it hard to find and compare all the rates of its reactions.
We have developed a tool~\cite{West:importer} that helps with this task of identifying the species in a detailed kinetic model~\cite{Lambert:2015vu}.
The tool was built using methods from the open-source Reaction Mechanism Generator (RMG)~\cite{Gao:2016dk} which predicts how identified molecules are expected to react.
Comparing this with how the \textsc{Chemkin} file says species reacts allows one to deduce which molecule corresponds to which species name.

We have used this tool to partially or fully import 74 detailed kinetic models gathered from the literature\cite{West:models}.
The 74 models include
20 from
\textit{Combustion and Flame} (2012--2015),
33 from
\textit{Proceedings of the Combustion Institute} (2013--2017),
and 21 models from other miscellaneous articles, provided by their authors, or downloaded from repository websites such as AramcoMech, USC-Mech, LLNL.gov, etc.
The full list is provided in the supplementary materials, and the models can be downloaded openly~\cite{West:models}.

\subsection{Alternate model generation}
The starting point for model generation was the LLNL comprehensive combustion model for the four butanol isomers by Sarathy et al.~\cite{Sarathy:2012fj}, which contains 418 species and $\sim$2343 reactions.
(Counting the number of reactions in a model is not without complications. The \textsc{Chemkin}-to-Cantera conversion script skips reactions with $k\equiv 0$, and converts explicit reverse reaction rates into two irreversible reactions in opposite directions.
Furthermore, RMG treats duplicates as a single reaction whose rate is given as the sum of multiple Arrhenius expressions, rather than as independent reactions.)

The large database of kinetic models\cite{West:models} was first filtered for only reactions containing at least one of the 418 species in the {original} model,
resulting in 55,775 instances of 13,618 rate expressions for 6,303 reactions occurring in 74 models.
To reduce the risk of introducing errors, kinetics that are represented as the sum of two Arrhenius expressions (i.e., duplicate reactions) were excluded from the analysis. This left 55,058 instances of 13,245 unique rates for 6,253 reactions.
Many of these reactions are the reverses of each other, but for the current analysis they were not merged or reversed, i.e., rates were only compared across models if the reactions were written in the same direction.
However, pressure-dependent and -independent reactions (e.g., \ce{A + B <=> C} and \ce{A + B\ ( +M) <=> C\ ( +M)}) were treated as alternative rate expressions for the same reaction.

For each reaction in the {original} model, the most common three rates occurring in the database were considered, but only if they appeared in at least two models.
For example, the reaction \ce{C4H3-n + OH <=> C4H2 + H2O} 
has a rate coefficient of $k= $ \SI{2e12}{\centi\meter^3\per\mole\per\second} in the {original} model and 22 others, 
but is \SI{2.5e12}{\centi\meter^3\per\mole\per\second} in three models, 
and \SI{1.5e13}{\centi\meter^3\per\mole\per\second} in two models. 
We also found the rate coefficient \SI{3.0e13}{\centi\meter^3\per\mole\per\second} in use, but this is the fourth-most common rate and only present in one model, 
so we excluded it from the current analysis.
All of these rates occur in models published in \textit{Proceedings of the Combustion Institute} or \textit{Combustion and Flame} since 2013.

We implemented the requirement that a rate expression has been seen ``in the wild'' at least twice to reduce the risk of possible errors made when importing the models, and to result in a reasonably conservative estimate of the impact of genuine discrepancies between ``accepted'' parameters, without being overly influenced by lone outliers.
{It should be noted, however, that a parameter appearing in many models does not indicate that it is correct.
This is not a popularity contest, and often the most accurately determined parameter is not the most commonly used.
Furthermore, a complete lack of discrepancy in the literature does not indicate a lack of uncertainty; often an uncertain estimate is adopted universally.}

In total there were 300 reactions with one alternative rate considered (besides the original), 471 with two alternatives, and 13 with three alternatives, totaling 1281 kinetic variants on the original model.
A similar process was undertaken for the thermochemical parameters.
These are provided in the \textsc{Chemkin} files in NASA polynomial form describing $\Delta H$, $S$, and $C_P(T)$.
When an alternative is considered, the full set of polynomials for that species are substituted.
Out of the 418 total species in the model, 65 species had one set of alternative thermodynamic parameters found in the database, 127 species had two alternatives, and 2 species had three alternatives, totaling 325 variations of thermodynamic parameters considered.

In total there are 1606 variants, when considered one at a time.

\subsection{Experimental data curation}

We collected experimental data from the literature that describe the autoignition in shock tubes of the four butanol isomers~\cite{Moss:2008bva,Stranic:2012jl,Bec:2014,Zhu:2014td} and converted these to the ChemKED standard~\cite{Weber:2017chemked}.
We did not consider measurements taken by Bec et al.~\cite{Bec:2014} and Zhu et al.~\cite{Zhu:2014td} using the constrained-reaction-volume method.
Weber and Niemeyer~\cite{Weber:2017chemked} describe the ChemKED (\textbf{Chem}ical \textbf{K}inetics \textbf{E}xperimental \textbf{D}ata) format in detail.
In brief, ChemKED is a human- and machine-readible, YAML-based file format for describing fundamental combustion measurements with sufficient information to simulate experimental data points.
Tables~\ref{T:nbutanol}--\ref{T:tbutanol} summarize these datasets for each isomer; in total, we considered 475 data points across 67 series (i.e., files).
All created ChemKED files are available openly~\cite{ChemKED-database}.

\begin{table}[htbp]
\centering
\begin{tabular}{@{}lccccc@{}}
\toprule
Study	& $P$ (atm) & $T$ (K) & $\phi$	& $\chi_{\ce{O2}}$ (\%) & \# points \\
\midrule
Moss et al.~\cite{Moss:2008bva}		& \numrange{0.95}{4.02} & \numrange{1196}{1711}& \numrange{0.25}{1.0} & \numrange{1.5}{24} & 44 \\
Stranic et al.~\cite{Stranic:2012jl}& \numrange{0.91}{45.12}& \numrange{1169}{1534}& \numrange{0.5}{1.0}  & \numrange{3}{4.5} & 37 \\
Bec et al.~\cite{Bec:2014}			& \numrange{14.7}{24.0} & \numrange{800}{1043} & \numrange{0.71}{3.34}& \numrange{6.12}{28.3} & 40 \\
Zhu et al.~\cite{Zhu:2014td}		& \numrange{15.1}{45.0} & \numrange{716}{1121} & \numrange{0.5}{2.0}  & \numrange{10.2}{40.6} & 37 \\
\bottomrule
\end{tabular}
\caption{Summary of experimental conditions for \textit{n}-butanol autoignition, where $\chi_{\ce{O2}}$ indicates
the molar percentage of \ce{O2} in the reactants.}
\label{T:nbutanol}
\end{table}

\begin{table}[htbp]
\centering
\begin{tabular}{@{}lccccc@{}}
\toprule
Study	& $P$ (atm) & $T$ (K) & $\phi$	& $\chi_{\ce{O2}}$ (\%) & \# points \\
\midrule
Moss et al.~\cite{Moss:2008bva}		& \numrange{1.2}{4.3} & \numrange{1196}{1654}& \numrange{0.25}{1.0} & \numrange{1.5}{24} & 36 \\
Stranic et al.~\cite{Stranic:2012jl}& \numrange{1.12}{43.55}& \numrange{1089}{1588}& \numrange{0.5}{1.0}  & \numrange{3}{4} & 41 \\
Bec et al.~\cite{Bec:2014}			& \numrange{15.3}{23.6} & \numrange{775}{1069} & \numrange{0.989}{1.033}& 20.3 & 11 \\
\bottomrule
\end{tabular}
\caption{Summary of experimental conditions for isobutanol autoignition, where $\chi_{\ce{O2}}$ indicates
the molar percentage of \ce{O2} in the reactants.}
\label{T:ibutanol}
\end{table}

\begin{table}[htbp]
\centering
\begin{tabular}{@{}lccccc@{}}
\toprule
Study	& $P$ (atm) & $T$ (K) & $\phi$	& $\chi_{\ce{O2}}$ (\%) & \# points \\
\midrule
Moss et al.~\cite{Moss:2008bva}		& \numrange{1.11}{4.01} & \numrange{1256}{1659}& \numrange{0.25}{1.0} & \numrange{1.5}{24} & 41 \\
Stranic et al.~\cite{Stranic:2012jl}& \numrange{1.16}{3.56}& \numrange{1169}{1534}& \numrange{0.5}{1.0}  & \numrange{3}{6} & 45 \\
Bec et al.~\cite{Bec:2014}			& \numrange{14.9}{20.7} & \numrange{828}{1062} & \numrange{1.04}{1.71}& \numrange{11}{20.3} & 13 \\
\bottomrule
\end{tabular}
\caption{Summary of experimental conditions for \textit{sec}-butanol autoignition, where $\chi_{\ce{O2}}$ indicates
the molar percentage of \ce{O2} in the reactants.}
\label{T:sbutanol}
\end{table}

\begin{table}[htbp]
\centering
\begin{tabular}{@{}llllll@{}}
\toprule
Study	& $P$ (atm) & $T$ (K) & $\phi$	& $\chi_{\ce{O2}}$ (\%) & \# points \\
\midrule
Moss et al.~\cite{Moss:2008bva}		& \numrange{1.10}{3.83} & \numrange{1263}{1825}& \numrange{0.25}{1.0} & 1.5 & 40 \\
Stranic et al.~\cite{Stranic:2012jl}& \numrange{0.99}{45.12}& \numrange{1169}{1534}& \numrange{0.5}{1.0}  & \numrange{3}{4} & 54 \\
Bec et al.~\cite{Bec:2014}			& \numrange{15.1}{21.0} & \numrange{955}{1095} & \numrange{1.04}{1.46}& \numrange{13.6}{20.3} & 7 \\
\bottomrule
\end{tabular}
\caption{Summary of experimental conditions for \textit{tert}-butanol autoignition, where $\chi_{\ce{O2}}$ indicates
the molar percentage of \ce{O2} in the reactants.}
\label{T:tbutanol}
\end{table}

\subsection{Model and data comparison}

We used the Python-based tool PyTeCK~\cite{Niemeyer:2016wf,Niemeyer:2016py} to automatically
test the performance of model variants across the full dataset for butanol isomer autoignition.
PyTeCK automatically creates Cantera~\cite{Cantera} simulations based on experimental measurements specified via ChemKED file, and runs these to find the simulated ignition delay for each condition.
It reports an overall performance metric for each model, mostly following that
used by Olm et al.~\cite{Olm:2014gn,Olm:2015ch} for hydrogen and syngas models.
The agreement between experimental and simulated ignition delay times is
quantified with the average error $E$ over the $N$ datasets given by
\begin{equation}
E = \frac{1}{N} \sum_{i=1}^N E_i \;,
\label{eq:E}
\end{equation}
where $E_i$ is the average error for the $i$th dataset
\begin{equation}
E_i = \frac{1}{N_i} \sum_{j=1}^{N_i} \left( \frac{ \log \tau_{ij}^{\text{exp}} - \log \tau_{ij}^{\text{sim}} }{ \sigma ( \log \tau_{ij}^{\text{exp}} ) } \right)^2 \;,
\end{equation}
$N_i$ is the number of data points in set $i$, $\tau_{ij}^{\text{exp}}$ and
$\tau_{ij}^{\text{sim}}$ are the experimental and simulated ignition delay times
for the $j$th data point in set $i$, respectively, and $\sigma$ is the standard deviation.
The $\log$ function indicates the natural logarithm.
Error quantities used the logarithm of ignition delay rather than the ignition
delay itself, following the practice of Olm et al.~\cite{Olm:2014gn,Olm:2015ch},
since the scatter in experimental results is proportional to the ignition delay value.

The standard deviation of an experimental dataset, $\sigma ( \log \tau_{ij}^{\text{exp}} )$
is determined by fitting a spline to the experimental ignition delay values with
respect to the variable changing in the dataset (typically temperature),
following the approach used by Olm et al.~\cite{Olm:2014gn,Olm:2015ch}.
This was implemented via the \texttt{interpolation.UnivariateSpline}
function~\cite{Walt:2011aa} of \texttt{SciPy}~\cite{scipy}.
A minimum value of $\sigma = 0.10$ was established, used to override any small calculated values.
This approach reduces the importance of experimental data sets with scatter, but does not attempt to identify or resolve systematic errors or inconsistencies between experimental data sets.

\subsection{Genetic algorithm to find extrema}
To assess the latitude possible by varying multiple parameters simultaneously, we used a genetic algorithm to find the fastest and slowest ignition delays at \SI{1000}{\kelvin}, \SI{10}{\bar}, and $\phi = 1.0$ in 4:96 \ce{O2}:\ce{N2}.
A genetic algorithm is a metaheuristic (i.e., an optimization method not guaranteed to find the global optimum, but likely to find a good one) inspired by the process of natural selection, using bio-inspired operators such as mutation, crossover, and selection.
Genetic algorithms are commonly used in high-dimensional optimization and search problems.
A genetic algorithm evolves a population of individuals toward better solutions; in this case, the genetic algorithm works to improve kinetic models towards slower or faster autoignition delay times.
Each individual has a set of properties (chromosome or genotype) which can be mutated and altered.
The chromosome or genotype for each kinetic model is a sequence of parameter choices, one for each reaction and species: 0 means use the original parameter, 1 means use the first alternative (if there is one), 2 means use the second, etc.
The chromosome has a length of 2777, which is equal to sum of number of species and reactions in the model.

\begin{figure}[htbp]
    \centering
    \begin{subfigure}{4.5 in}
        \centering
        \includegraphics[width=\textwidth]{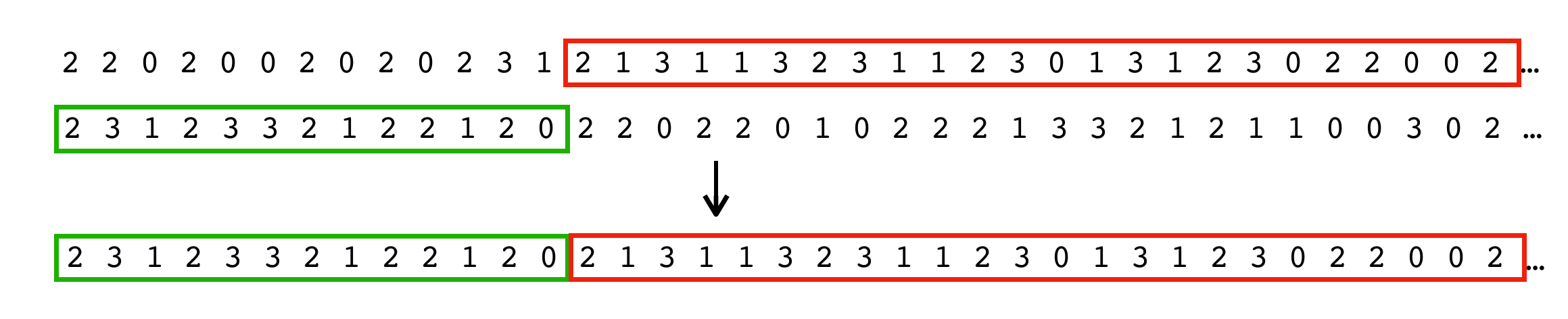}
        \caption{Crossover event}
        \label{fig:GA_crossover}
    \end{subfigure}
    \\
    \begin{subfigure}{4.5 in}
        \centering
        \includegraphics[width=\textwidth]{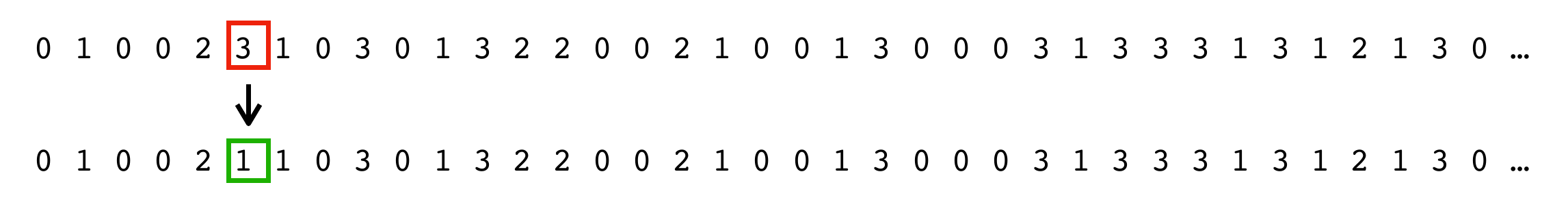}
        \caption{Mutation event}
        \label{fig:GA_mutation}
    \end{subfigure}
    \caption{Demonstration of genetic algorithm events}
    \label{fig:geneticalgorithm}
\end{figure}

In every generation the chromosomes are cloned, mutated, or crossed over with other individuals to generate offspring, as illustrated in Fig.~\ref{fig:geneticalgorithm}.
The crossover event selects two genomes as parents then randomly chooses a point along the genome to switch from one to the other, creating an offspring.
The mutation event picks a random genome, picks a random parameter (gene) with more than one option, and randomly selects from the available options to create an offspring.

After the first generation the randomly generated initial population and offspring are allowed to evolve based on a selection criteria or fitness function.
To find the extrema we used a fitness function of ignition delay time (defined as time of maximum $dT/dt$) for a constant-volume adiabatic batch reactor starting at \SI{1000}{\kelvin}, \SI{10}{\bar}, $\phi$=1, in a 4:96 mix of \ce{O2} and \ce{N2}, and minimized or maximized this function.
The evolution was run with different combinations of mutation, crossover, and selection parameters to find the best solution.

We used the Python-based tool DEAP (Distributed Evolutionary Algorithms in Python)~\cite{DEAP_JMLR2012} to run the genetic algorithm, and Cantera~\cite{Cantera} to evaluate the ignition delay times.
The initial population contains 100 randomly generated individuals;
the crossover probability between two individuals was set to 0.8,
and the mutation probability for one individual was set to 0.2.
The number of individuals to select for the next generation was $\mu$=100 and the number of offspring to produce at each generation was $\lambda$=200.
We used the genetic algorithm called $(\mu + \lambda)$, meaning the next generation of individuals are selected from the pool of both parent and offspring populations.
Selection was done using tournaments of size three.
There was a 70\% chance that an offspring would be generated by single-point crossover from two parents (Fig.~\ref{fig:GA_crossover}),
a 20\% chance that an offspring would be generated by mutating a single random gene from one parent (Fig.~\ref{fig:GA_mutation}),
and a 10\% chance that an offspring would be generated by cloning a parent.

The optimization was run for 500 generations when varying just kinetics or just thermodynamics, but when varying all parameters the optimization had not converged after 500 and so it was run for 1000 generations.
At the end of the optimization, the fastest and slowest models (at \SI{1000}{\kelvin} and \SI{10}{\bar}) were then run through the full PyTeCK comparison against all the experimental data.

{To further explore the extremes that could be reached while allowing all parameters to be changed,
we performed additional optimizations with different objectives:
to find the extrema (maximum and minimum) of ignition delay at \SI{1500}{\kelvin} and \SI{43}{\atm},
and to find the extrema of the average slope in the ignition delay curve between \SI{900}{\kelvin} and \SI{1500}{\kelvin} at \SI{43}{\atm}, i.e., to maximize and minimize $\left(\log\tau_{(\SI{900}{\kelvin}, \SI{43}{\atm})} - \log\tau_{(\SI{1500}{\kelvin}, \SI{43}{\atm})}\right).$}

\FloatBarrier
\section{Results and discussion}

\begin{figure}[htbp]
\centering
\includegraphics[width=4.5in]{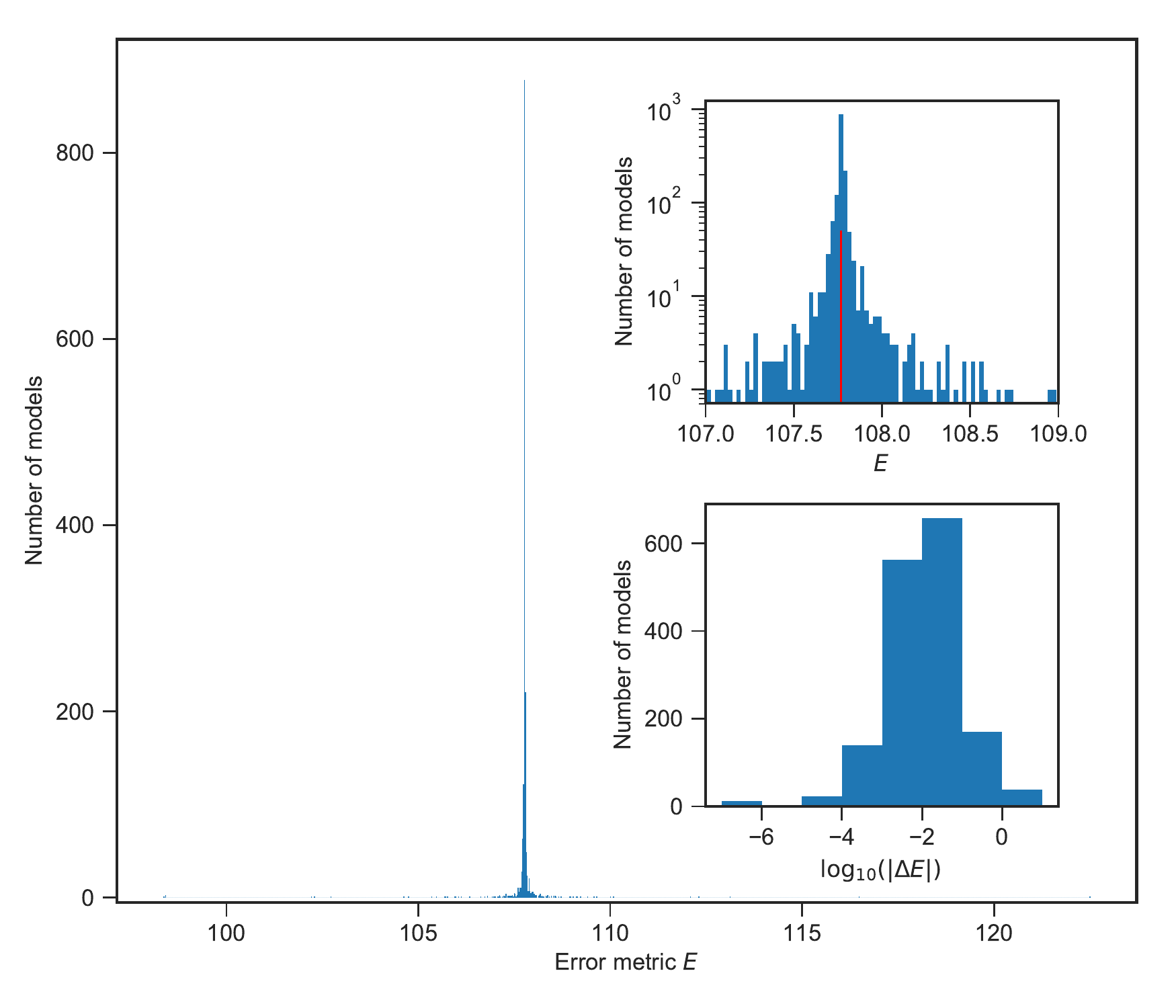}
\caption{Histograms of error metric $E$ from 1606 models, values of which range from 98.36 to 122.51.
Most variations change $E$ by less than 0.01 from the original value of $E_0=107.766$ (the red line), as shown by the two inset plots: the first with a logarithmic $y$ axis and a zoomed-in $x$ axis, the second showing a histogram of $\log_{10}\left(\left| E-E_0 \right|\right)$. From \cite{supplementary}}
\label{fig:histograms}
\end{figure}

The original model by Sarathy et al.~\cite{Sarathy:2012fj} has an overall error metric (see Eq.~\eqref{eq:E}) of $E$ = 107.76669{ representing the error over all 475 data points for all experimental conditions (Tables 1--4)}.
Two-thirds of the 1606 individual variations change this value by less than 0.01 and half of them by less than 0.001.
However, some variants decreased the error by as much as $-$9.4 to 98.36 or increased it by +14.7 to 122.51.
These outliers lead to a sharp histogram of $E$ values when the $x$-axis is scaled to show the full range, as shown in Fig.~\ref{fig:histograms}; the inset plots show greater detail.

Table~\ref{tab:influencers} lists the 25 substitutions that most change the error.
Five of these 25 variants are thermodynamic parameters and 20 are kinetics parameters.
Table~\ref{tab:thermo} shows details of the thermodynamic changes.
Although often a date is included, most published models omit or strip comments from their thermochemistry data files, making it difficult to establish where the parameters came from.
In most cases the thermochemical parameters in the original model match those in
 Burcat's database with updates from the Active Thermochemical Tables \cite{burcat2005third}
or some subsequent version of the extended database now maintained online \cite{burcat2017extended}.
Many of the variants commonly in use were estimated using the THERM software\cite{Ritter:2002bf} which is based on Benson's group additivity method\cite{Benson:1976}.

\begin{table}[htbp]
\caption{Most influential parameter substitutions on overall error metric, listed in descending order of $\left|\Delta E\right|$}
\centering
\begin{tabular}{ l  r  c r r  }
\toprule
Type& Reaction or & Variant  & $E$ & $\Delta E$ \\
    &  Species No. &  No.    &     &  \\
\midrule
Thermochemistry   & 190   & 1 & 122.514 & \num{+14.747} \\
Kinetics          & 293   & 2 & 98.359 & \num{-9.408} \\
Kinetics          & 187   & 2 & 98.420 & \num{-9.347} \\
Kinetics          & 59    & 1 & 98.422 & \num{-9.344} \\
Thermochemistry   & 224   & 1 & 98.430 & \num{-9.337} \\
Thermochemistry   & 190   & 2 & 116.494 & \num{+8.728} \\
Kinetics          & 187   & 1 & 102.222 & \num{-5.545} \\
Kinetics          & 1440  & 1 & 102.276 & \num{-5.490} \\
Kinetics          & 61    & 1 & 102.319 & \num{-5.447} \\
Kinetics          & 291   & 1 & 113.119 & \num{+5.353} \\
Kinetics          & 180   & 1 & 102.707 & \num{-5.059} \\
Kinetics          & 272   & 1 & 112.315 & \num{+4.548} \\
Kinetics          & 272   & 2 & 112.085 & \num{+4.318} \\
Kinetics          & 391   & 2 & 104.632 & \num{-3.134} \\
Thermochemistry   & 90    & 1 & 104.744 & \num{-3.023} \\
Kinetics          & 1441  & 1 & 105.343 & \num{-2.424} \\
Kinetics          & 535   & 1 & 110.085 & \num{+2.318} \\
Kinetics          & 391   & 1 & 105.461 & \num{-2.305} \\
Kinetics          & 535   & 2 & 110.008 & \num{+2.241} \\
Kinetics          & 189   & 2 & 105.700 & \num{-2.067} \\
Kinetics          & 1267  & 1 & 105.714 & \num{-2.053} \\
Kinetics          & 1441  & 2 & 105.767 & \num{-1.999} \\
Thermochemistry   & 107   & 1 & 109.659 & \num{+1.892} \\
Kinetics          & 168   & 2 & 105.950 & \num{-1.817} \\
Kinetics          & 321   & 2 & 109.583 & \num{+1.816} \\
\bottomrule
\end{tabular}
\label{tab:influencers}
\end{table}





\begin{table}[htbp]
\renewcommand{\arraystretch}{1.5}
\linespread{0.6}\selectfont\centering
\caption{Thermochemical parameters for which substitutions most impact $E$.}
\label{tab:thermo}

\makebox[\textwidth][c]{ 
\begin{tabular}{@{}c l >{\raggedright\footnotesize}p{1.5in} >{\raggedright\footnotesize}p{1.1in} r @{}}
\toprule
No.\ &  Molecule & {\normalsize Original source} & {\normalsize Variant source} & $\Delta E$ \\
\midrule
190(Var1) &
\ce{ iC4H8 }
&
Burcat (2009)\cite{burcat2017extended}
&
THERM\cite{Ritter:2002bf}
&
+14.747
\\

224 &
\ce{ iC3H5OH }
&
Unknown; possibly THERM
&
THERM\cite{Ritter:2002bf}
&
--9.337
\\

190(Var2) &
\ce{ iC4H8 }
&
Burcat (2009)\cite{burcat2017extended}
&
Burcat (1983)\cite{burcat2005third}
&
+8.728
\\

90 &
\ce{ C3H6 }
&
Burcat (2000)\cite{burcat2005third}
&
Unknown (1986)
&
--3.023
\\

107 &
\ce{ nC3H7O2  }
&
Burcat (2010)\cite{burcat2017extended}
&
THERM \cite{Ritter:2002bf}
&
+1.892
\\
\bottomrule
\end{tabular}
}

\end{table}

\FloatBarrier
Table~\ref{tab:reactions-decreasing} lists the 10 reactions corresponding to the 12 kinetic substitutions that most decrease the overall error metric.
Table~\ref{tab:effects-decreasing} shows details of those substitutions: the $\Delta E$, the parameter values, and the source of those values, as best as we can determine.
Although many researchers follow the helpful practice of including comments in their \textsc{Chemkin} files indicating where they think a value came from, most often these point to another model which in turn got it from somewhere else.
Tables~\ref{tab:reactions-increasing} and~\ref{tab:effects-increasing} show the same information corresponding to the eight kinetic substitutions with the largest effect of \textit{increasing} the overall error metric.
For the top influencer in each list (reactions 293 and 291) we discuss the source of the parameters in greater depth{ below}.
However, the aims of this paper are to demonstrate the new tools and to investigate the impact of discrepancies in the literature, not to resolve the discrepancies or to create another model for butanol, so we restrict this analysis to two parameters in addition to the notes in Tables~\ref{tab:effects-decreasing} and~\ref{tab:effects-increasing}.

\begin{table}[htbp]
\centering
\caption{Reactions for which substitutions have the greatest effect of reducing error metric $E$ (see Table~\ref{tab:effects-decreasing} for more details).}
\label{tab:reactions-decreasing}
\begin{tabular}{@{}c r @{\ce{<=>}} l @{}}
\toprule
No. & \multicolumn{2}{c}{ Reaction} \\
\midrule
293 & \ce{HCCO +O2 }&\ce{ CO2 + CO + H} \\
187 & \ce{CH + H2O }&\ce{ H + CH2O} \\
59 & \ce{C2H + O }&\ce{ CO + CH } \\
1440 & \ce{iC4H8 }&\ce{ C3H5 + CH3} \\
61 & \ce{C2H + O2 }&\ce{ CO2 + CH } \\
180 & \ce{CH + O2 }&\ce{ CHO + O } \\
391 & \ce{CH3COCH3 }&\ce{ CH3CO + CH3 } \\
1441 & \ce{iC4H8 }&\ce{ iC4H7 + H} \\
189 & \ce{CH3 + CH3 }&\ce{ C2H6} \\
1267 & \ce{iC4H8 + H}&\ce{ iC4H9} \\

\bottomrule
\end{tabular}
\end{table}

\FloatBarrier
\paragraph{Reaction 293}
The kinetics substitution which would improve the model performance the most (decrease its error metric $E$) is reaction 293:
\begin{equation*}
\ce{HCCO +O2 <=> CO2 + CO + H} \;.
\end{equation*}
The original model used the rate
$k=\num{4.78e12}\left(\frac{T}{\SI{1}{\kelvin}}\right)^{-0.142}\exp\left(\frac{-\SI{1.15}{\kilo\calorie\per\mole}}{R T}\right)$.
The source of this rate is
Klippenstein, Miller, and Harding (2002) who used electronic structure theory, RRKM theory, and master equation and trajectory simulations to solve a mystery of prompt \ce{CO2} formation, and provided rate coefficients for \SIrange{300}{2500}{\kelvin}. They used QCISD(T) and MP2 energies from B3LYP geometries, then lowered the barrier by \SI{3.2}{\kilo\calorie\per\mole} to match room-temperature experimental data, ending up with a good agreement with experiments.

The alternative rate for substitution is
$k=\num{6.0e12}\exp\left(\frac{-\SI{0.859}{\kilo\calorie\per\mole}}{R T}\right)$.
One model 
attributes the rate to
Baulch et al.~(1992)
but the numbers \citep[p.713]{Baulch:1992uq} do not quite match---the
model is about four times faster.
Baulch et al.\ provide a rate for \SIrange{300}{550}{\kelvin} 
but warn that ``Kinetic data on this reaction are very limited, and no products have been suggested'' and assigned a reliability of $\Delta \log k = \pm 0.7${ in that temperature range}.
Another model 
notes that the source was Baulch et al.~\cite{Baulch:1992uq} modified by Zeuch (2003)~\cite{Zeuch:2003ua}.
Zeuch's PhD thesis contains the numbers in use~\citep[p.210]{Zeuch:2003ua},
and explains that the products and rate constant were changed from that of Baulch et al.\
to better fit some flame speeds.
Without attempting to judge if the change was an improvement, we note that some researchers probably think they are using Baulch numbers in their models, when they are not.

Although the electronic structure methods available at the time do not match the accuracy of those in use today, it is {in our view }most likely that the 2002 Klippenstein et al.\ rate expression is closer to the truth than the rate based loosely on ``very limited'' data with ``uncertain experimental conditions''.
Yet, replacing the former with the latter is the single substitution which most \textit{decreases} the overall error metric for this model ($E \downarrow 9.408$).
This is a case where a focus on getting closer to the data would take one further from the truth, and we do not endorse such a substitution.

\begin{table}[htbp]
\makebox[\textwidth][c]{ 
\renewcommand{\arraystretch}{1.6}
\linespread{0.6}\selectfont\centering
\begin{threeparttable}[t]
\caption{Kinetic substitutions that have the greatest effect of decreasing the error metric $E$ (see Table~\ref{tab:reactions-decreasing})}
\label{tab:effects-decreasing}
\begin{tabular}{@{}c  l c c c >{\footnotesize}p{2.5in} c @{}}
\toprule
No. & &   $A$  &  $n$ & $E_a$ & Source & $\Delta E$\\
& & (\si{\centi\meter^3},\si{\mole},\si{\second}) & &  (\si{\kilo\calorie\per\mole}) &  & \\
\midrule

293 & Orig.\ & \num{ 4.78E+12 } & \num{-0.142} & 1.15  &
Klippenstein (2002)~\cite{Klippenstein:2002hz}  &
--
\\ 
  & Repl.\ & \num{ 6.00E+12} &  0.00 &  0.859 &
 Baulch (1992)~\cite{Baulch:1992uq}; modified by Zeuch (2002)~\cite{Zeuch:2003ua}. See text.
&
\num{-9.408}
  \\
\midrule

187 & Orig.\ & \num{ 1.71E+13 } & 0.0 & \num{-0.775} &
GRI Mech version 2.1; originally from Baulch et al.\ (1992)~\cite{Baulch:1992uq} but increased by factor of 3 during model optimization.  & 
--
\\ 
  & Var.~1 & \num{ 5.71E+12 } &  0.0 & \num{-0.755} &
GRI Mech 3.0~\cite{GRI3.0}; originally from Baulch et al.\ (1992)~\cite{Baulch:1992uq} and unchanged during model optimization. &
\num{-5.545}
\\ 
  & Var.~2 & \num{ 1.77E+16 } &  \num{-1.22} & 0.0238 &
 Bergeat et al.\ (2009)~\cite{Bergeat:2009je}&
\num{-9.347}
  \\

\midrule

59 & Orig.\ & \num{ 6.2E+12 } & 0.0 & 0.0 &
For excited \ce{CH^*} formation. (Ground state formation not included in model.) Attributed in some other models to Kathrotia et al.\ (2010), but in fact originating from Smith et al.\ (2002)\cite{Smith:2002wh} (who revised it in 2005\cite{Smith:2005fl} to \num{2.5e12}).&
--
\\ 
  & Var.\ & \num{ 5.0E+13 } &  0.0 & 0.0 &
For ground state \ce{CH} formation.
(Some models have both Orig.\ and Var.\ reactions.)
GRI Mech 3.0~\cite{GRI3.0}; originally from Browne et al.~(1969) with note ``no experiments''.
&
\num{-9.344}
  \\
\midrule

1440 & Orig.\ & \num{ 1.92E+66 } & \num{-14.22} & 1281  &
``HENRY?''\tnote{c}&
--
\\ 
  & Var.\ & \num{3.30E+21} &  \num{-1.2} & 97.72 &
 Schenk et al.~(2013), probably using a rate for \ce{C3H6 -> C2H3 + CH3}.
&
\num{-5.490}
  \\
\midrule

61 & Orig.\ & \num{ 2.17E+10 } & 0.0 & 0.0 &
Hwang et al.~(1987)~\citep{HwangS.M.1987Izeo}  &
--
\\ 
  & Var.\ & \num{ 4.50E+15 } &  0.0 & 25.0 &
JetSurF 2.0~\cite{JetSurF2.0}&
\num{-5.447}
  \\
\midrule

180 & Orig.\ & \num{ 3.30E+13 } & 0.0 & 0.0 &
Baulch et al.\ (1992)~\cite{Baulch:1992uq} (as used in GRI Mech 2.1)&
--
\\ 
  & Var.\ & \num{ 6.71E+13 } &  0.0 & 0.0 &
GRI Mech 3.0~\cite{GRI3.0}; updated in 3.0 release: average of two rates from 1996 and 1997, reduced 21\% during model optimization.&
\num{-5.059}
  \\
\midrule

391 & Orig.\tnote{a} & \num{ 7.108e21 } & \num{-1.570} & 84.68 &
``HENRY''\tnote{c} &
--
\\ 
  & Var.~1\tnote{b} & \num{ 9.40E+28 } &  \num{-3.669} &  89.02 &
Saxena et al.\ (2009)~\citep{Saxena:2009hh}&
\num{-3.134}
  \\
  & Var.~2  & \num{ 1.21E+23 } &  \num{-1.99} & 83.95 &
Unknown; used in ``MB-Farooq'' and ``MB-Fisher''&
\num{-2.305}
  \\

\midrule


1441 & Orig.\ &  \num{3.07e55}  & \num{-11.49} & 114.3 &
``HENRY?''\tnote{c} &
--
\\ 
  & Var.\tnote{b}&  \num{4.73e60} & \num{-1.266} & 114.404&
  AramcoMech~2.0.
Based on $k_{\infty}$ from \ce{C3H5 + H ({+}M) <=> C3H6 ({+}M)} (original source unknown) with modifications. &
\num{-2.424}
  \\
\midrule


189 & Orig.\tnote{a} & \num{ 2.277e+15 } & \num{-0.69} & 0.174 &
Wang et al.~(2003)\cite{Wang:2003hl}&
--
\\ 
  & Var.\tnote{a} & \num{2.12e16 } &  \num{-0.97} & 0.620 &
 GRI-Mech 1.1, 1.2, and 2.1 (but not 3.0); originally from Stewart et al.\ (1989) Combust.\ Flame 75, 25
 &
\num{-2.067}
  \\
\midrule

1267 & Orig.\ & \num{ 6.25E+11 } & 0.51 & 2.62 &
Curran (2006)\cite{Curran:2006gb}  &
--
\\ 
  & Var.\tnote{a} & \num{ 1.33E+13 } &  0.0 &  3.26 &
Value for \ce{C3H6 + H <=> nC3H7} from USC-Mech II &
\num{-2.053}
  \\

\bottomrule
\end{tabular}
 \vspace{0.5ex}
 \begin{tablenotes}
  \item [a] Troe fall-off reaction. High pressure limit $k_\infty$ reported here.
   \item [b] pressure-dependent (PLOG) expression. Highest available pressure reported here.
   \item [c] presumably a reference to model co-author Henry Curran
 \end{tablenotes}
\end{threeparttable}
} 
\label{tab:reactions}
\end{table}

\begin{table}[htbp]
\centering
\caption{Reactions for which substitutions have the greatest effect of increasing error metric $E$ (see Table~\ref{tab:effects-increasing})}
\label{tab:reactions-increasing}
\begin{tabular}{@{}c r @{\ce{<=>}} l @{}}
\toprule
No. & \multicolumn{2}{c}{ Reaction} \\
\midrule
291 & \ce{HCCO + H }&\ce{ CH2 + CO } \\
272& \ce{CH2CHO + O2 }&\ce{ CH2CO + OOH} \\
535 & \ce{nC3H7 + O2 }&\ce{ C3H6 + OOH } \\
321 & \ce{C2H4 + CH3 }&\ce{ C2H3 + CH4 } \\
145 & \ce{CH3 + HO2 }&\ce{ CH3O + OH} \\
\bottomrule
\end{tabular}
\end{table}

\paragraph{Reaction 291}
The kinetics substitution which would worsen the model performance the most (increase its error metric $E$) is reaction 291:
\begin{equation*}
\ce{HCCO + H <=> ^1CH2 + CO} \;.
\end{equation*}
The original model used the rate
$k=\SI{1e14}{\centi\meter^3\per\mole\per\second}$, attributed to GRI-Mech 3.0~\cite{GRI3.0} which reports the source as Miller and Bowman (1989)~\cite{Miller:1989}
whilst noting ``No measurements; estimated.''

The alternative rate 
 $k=\SI{1.1e13}{\centi\meter^3\per\mole\per\second}$, is nine times slower,
was found in three models,
and is attributed to
Healy et al.\ (2008)~\cite{Healy2008Methanepropane}.
But that model took it unmodified
from Petersen et al.\ (2007)\cite{Petersen:2007bz},
who in turn report ``the methane/ethane system is based on that published by Fischer et al.\ (2000)~\cite{Fischer:2000}''.

Although Petersen et al.~\cite{Petersen:2007bz} state
``modifications have been made to some of the methane chemistry''
this reaction was not one of ``the more significant changes discussed here,'' possibly due to page limits in the Symposium proceedings.
However, the reported source uses a value 10 times higher, $k=\SI{1.1e14}{\centi\meter^3\per\mole\per\second}$~\cite[R97]{Fischer:2000},  
citing Miller et al.\ (1992)~\cite{Miller:1992hw}
who discuss the sensitivity of the pathway for soot formation and use $k=\SI{1.0e14}{\centi\meter^3\per\mole\per\second}$~\citep[R126]{Miller:1992hw},
referring to Miller et al.\ (1991)~\cite{Miller:1991fm} for model details.
That paper, in turn, notes it is the most important step for removal of \ce{HCCO} in rich flames, and that the rate used is consistent with those determined by Peeters and coworkers (1985, 1986)~\cite{Vinckier:1985bp,Peeters:1986cs}.
These {papers }reveal (a) the \num{1.1e14} value was in fact for the reaction
\ce{HCCO + O <=> 2 CO + H}
and the reaction in question is 1.4 times faster,
and (b) the value refers to \SI{535}{\kelvin} and there is an activation energy of $0.6 \pm 0.3$~kcal/mol in the range \SIrange{285}{535}{\kelvin}~\cite{Vinckier:1985bp}.
Furthermore, the rate $(1.55 \pm 0.45)\times\SI{e14}{\centi\meter^3\per\mole\per\second}$ at \SI{535}{\kelvin} was for \ce{HCOO + H -> products} and the branching ratio was not accurately known~\cite{Peeters:1986cs}.

In summary, the rate was determined in 1985, rounded down and made temperature-independent in 1991, copied in 1992, increased 10\% in 2000,
reduced by a factor of 10 in 2007, copied in 2008, and then used several times since.
The NIST Kinetics database~\cite{Manion:2017uv} gives two other values of \num{1e14} and \SI{1.5e14}{\centi\meter^3\per\mole\per\second}.
Little justification is given for the \SI{1.1e13}{\centi\meter^3\per\mole\per\second} value, so probably the models that copy it could reconsider its use.
In any case, adopting this rate would \textit{worsen} the performance of the current butanol model, i.e., increase its error metric $E$ by $+5.353$.

\begin{table}[htbp]
\makebox[\textwidth][c]{ 
\renewcommand{\arraystretch}{1.5}
\linespread{0.6}\selectfont\centering
\begin{threeparttable}[t]
\caption{Kinetic substitutions that increase the error metric $E$.}
\label{tab:effects-increasing}
\begin{tabular}{@{}c  l c c c >{\footnotesize}p{2.5in} c @{}}
\toprule
No.\ & &   $A$  &  $n$ & $E_a$ & Source & $\Delta E$\\
& & (\si{\centi\meter^3},\si{\mole},\si{\second}) & &  (\si{\kilo\calorie\per\mole}) &  & \\
\midrule

291 & Orig.\ & \num{ 1.0E+14 } & 0.0 & 0.0 &
GRI-Mech 3.0~\cite{GRI3.0},
originally from Miller and Bowman (1989)~\cite{Miller:1989}&
--
\\ 
  & Var.\ & \num{ 1.1E+13 } &  0.0 &  0.0 &
See discussion in main text for complicated genealogy~\cite{Healy2008Methanepropane,Petersen:2007bz,Fischer:2000,Miller:1992hw,Miller:1991fm,Vinckier:1985bp,Peeters:1986cs,Manion:2017uv}
&
+5.353
  \\
\midrule

272 & Orig.\tnote{a}& \num{ 7.05e7} & 1.63 & 25.29 &
Lee and Bozzelli (2003)~\cite{Lee:2003kt}&
--
\\ 
  & Var.~1 & \num{ 1.40E+11 } &  0.0 & 0.0 &
Baulch et al.\ (1992)~\cite{Baulch:1992uq}&
+4.548
\\
	& Var.~2  & \num{ 5E+11 } &  0.0 & 3.0 &
Unknown, but common author to three models using it is T.\ Faravelli at Politecnico di Milano &
%
+4.318
  \\
\midrule


535 & Orig.\ & \num{ 3.00E-19} & 0.0 & 3.00 &
Unknown. Probably Curran (1998)\cite{Curran:1998bx} with $A=\num{3E+11}$ reduced by $10^{30}$ around 2007. 
&
--
\\ 
  & Var.~1 & \num{ 9.00E+10 } &  0.0 &  0.0 &
Tsang (1988)~\cite{Tsang:1988kd}&
+2.318
  \\
  & Var.~2 & \num{ 3.7E+16 } & \num{-1.63} & 3.42 &
 DeSain, Klippenstein, et al.\ (2003)\cite{DeSain:2003gj} &
+2.241
  \\
\midrule


321 & Orig.\ & \num{ 6.62e0 } & 3.7 & 9.5 &
 Tsang and Hampson (1986)~\citep[p.1191]{Tsang:1986ek} estimated by Tsang in 1984 with uncertainty factor of 2 &
--
\\ 
  & Var.~1  & \num{ 2.27E+5 } &  2.0 & 9.2 &
GRI-Mech 3.0~\cite{GRI3.0} created by fitting to Kerr and Parsonage (1976)~\cite{kerr1976evaluated}&
+1.447
\\ 
  & Var.~2  & \num{ 6.3E+11 } &  0.0 & 16.0 &
Ahonkhai et al.\ (1989)~\cite{Ahonkhai:1989ha} &
+1.816
  \\
\midrule

145 & Orig.\ & \num{1.00e12} & 0.269 & -0.688 &
Jasper, Klippenstein, Harding (2009)~\citep{Jasper:2009ho} &
--
\\ 
  & Var.\ & \num{ 1.80e13 } &  0.0 & 0.0 &
 Baulch et al.\ (1992)~\cite{Baulch:1992uq}&
+1.479
  \\

 \bottomrule
 \end{tabular}
 \vspace{0.5ex}
 \begin{tablenotes}
  \item [a] Pressure-dependent (PLOG) expression; highest available pressure reported here.
 \end{tablenotes}
\end{threeparttable}
} 
\end{table}

\FloatBarrier

\subsection{Optimization using genetic algorithms}

The kinetic models were evolved using genetic algorithms to give faster or slower ignition delay times at \SI{1000}{\kelvin}, \SI{10}{\bar}, $\phi = 1$ and 4\% \ce{O2} in \ce{N2} bath gas.
For each objective  (maximizing or minimizing the ignition delay) the evolution was run three times: while substituting rates only, thermodynamic parameters only, and rates plus thermodynamic parameters simultaneously.
The ignition delay time $\tau$ was calculated for every model at each generation.

\begin{figure}[htbp]
\centering
\includegraphics[width=80 mm]{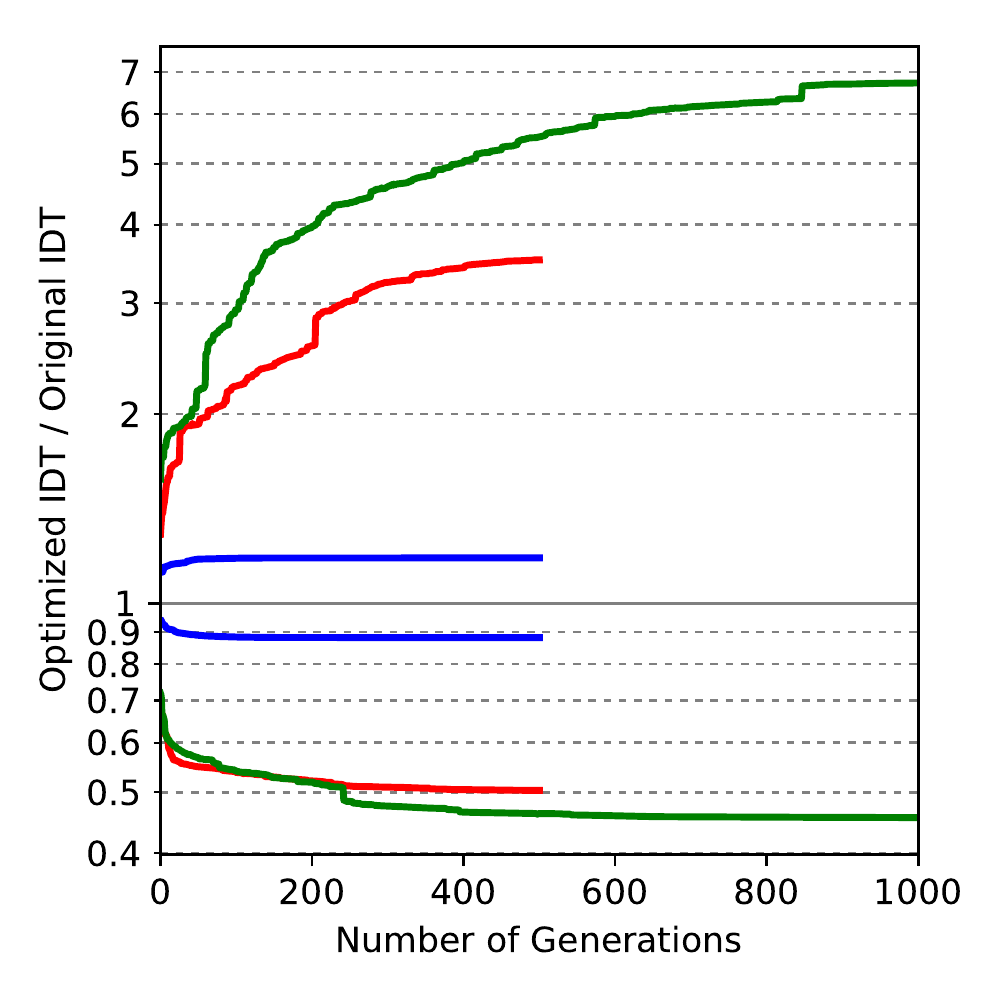}
\caption{Progress of the genetic algorithm to maximize and minimize ignition delay times (at \SI{10}{\bar}, \SI{1000}{\kelvin}, $\phi=1$ in 4:96 \ce{O2} and \ce{N2}). Blue: varying thermochemistry only (500 generations); red: varying kinetics only (500 generations); green: varying all parameters (1000 generations).}
\label{fig:optimization-progress}
\end{figure}

Figure~\ref{fig:optimization-progress} shows the average $\tau$ of the population for every generation, normalized using $\tau$ of the original model, for six runs of the genetic algorithm.
The climbing curves show evolutions with the fitness function designed to maximize ignition delay time, and the falling curves were minimizing ignition delay time.
In the blue curves only the thermochemical parameters were allowed to vary, in the red curves only the kinetics, and in the green curves all of the alternatives were allowed.
The latter cases were run for 1000 generations because the maximization curve was still climbing steadily after 500 generations.
In total 718,574 models were tested (about 180,000 for each of thermodynamic and kinetics, and 359,000 for the combined runs).

The original ignition delay time for these conditions was \SI{12.940}{\milli\second}{ at \SI{1000}{\kelvin} and \SI{10} bar}.
By varying all parameters, the model can be slowed by a factor of almost seven, to \SI{86.996}{\milli\second}, or sped up by a factor of about two, to \SI{5.899}{\milli\second}.
Most of the latitude comes from the kinetic parameters.
When the fastest and slowest models (at \SI{1000}{\kelvin} and \SI{10}{\bar}) were then run by PyTeCK with all the experimental data, the error metric $E$ increased +31.33 to 139.10 and  +66.74 to 174.51, respectively.

\begin{figure}[htbp]
\centering
\includegraphics[width=80 mm]{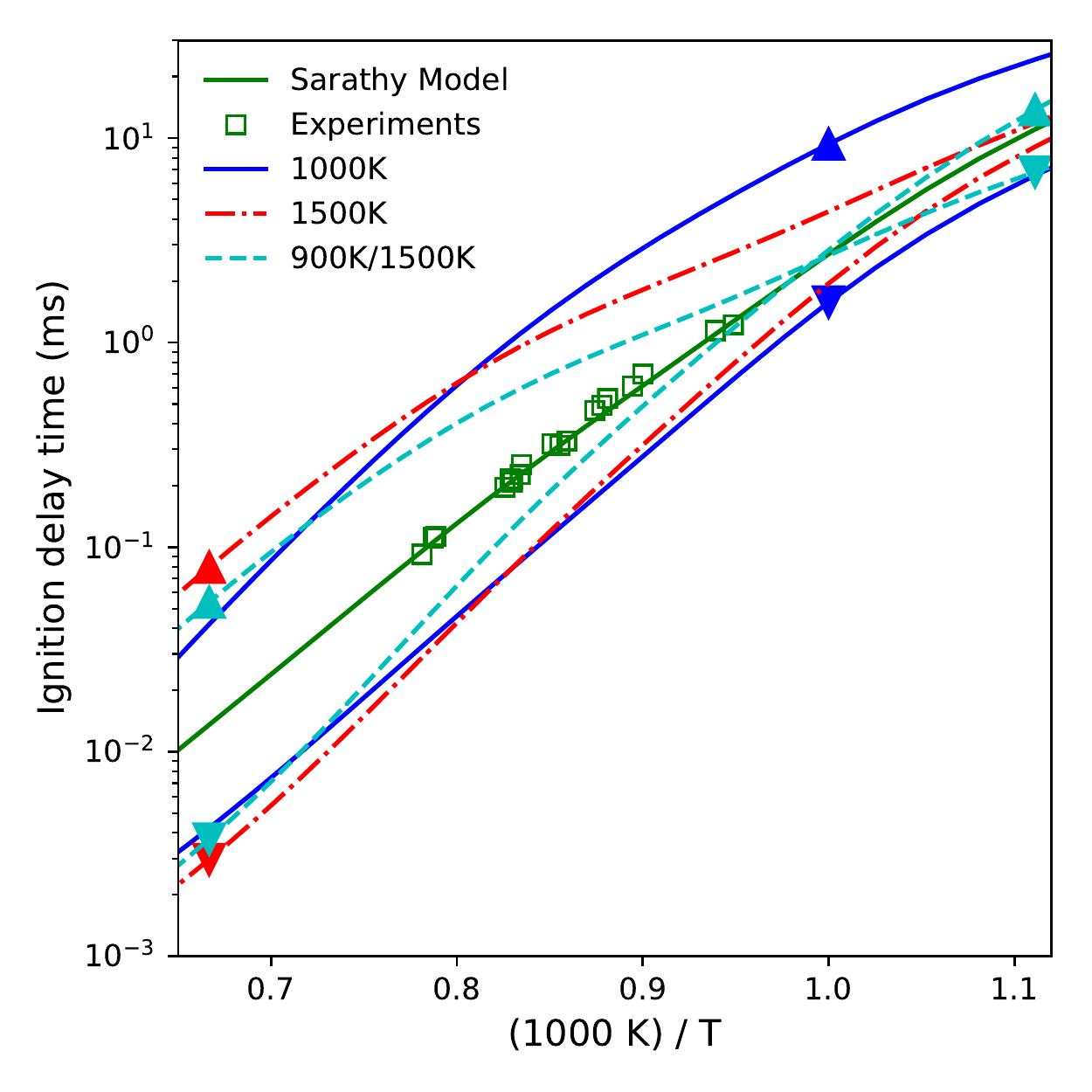}
\caption{
Ignition delay curves from the original model\cite{Sarathy:2012fj} (green) and the experimental data used in its construction\cite{Stranic:2012jl} (green squares), and the results of genetic algorithm with various objectives that were maximized or minimized:   $\tau_{(\SI{1000}{\kelvin}, \SI{10}{\bar})}$ (blue solid),  $\tau_{(\SI{1500}{\kelvin}, \SI{43}{\atm})}$ (red dot-dash), and $\left(\log\tau_{(\SI{900}{\kelvin}, \SI{43}{\atm})}-\log\tau_{(\SI{1500}{\kelvin}, \SI{43}{\atm})}\right)$ (teal dashed). The colored arrowheads indicate the optimization objectives. From \cite{supplementary}.
}
\label{fig:ignition-delays}
\end{figure}

{We performed four further optimizations (each for 1000 generations, allowing all parameters to change) with different objectives:
to find the maximum and minimum ignition delay at \SI{1500}{\kelvin} and \SI{43}{\atm},
and to find the maximum and minimum of the average slope in the ignition delay curve between \SI{900}{\kelvin} and \SI{1500}{\kelvin} at \SI{43}{\atm}, i.e., to maximize and minimize $\left(\log\tau_{(\SI{900}{\kelvin}, \SI{43}{\atm})} - \log\tau_{(\SI{1500}{\kelvin}, \SI{43}{\atm})}\right)$.
The models resulting from these optimizations were then used to calculate ignition delay curves  at \SI{43}{\atm}, $\phi = 1$ in 4:96 \ce{O2} and \ce{Ar},
which Figure~\ref{fig:ignition-delays} shows along with the original model~\cite{Sarathy:2012fj} and the experimental data from Stranic et al.~\cite{Stranic:2012jl} used in the construction/validation of the original model.
This further illustrates the wide range of results that could be achieved by indiscriminately picking from recently published model parameters.}

\section{Conclusions}
We present powerful new tools assembled into a novel workflow to assess the impact of discrepancies amongst kinetic rate expressions and thermochemical data in common use \cite{supplementary}.
Most discrepancies minimally affect a model's overall performance in predicting ignition delays, although some have a significant effect.
There are so many discrepancies to choose from that by cherry-picking parameters, each with defensible arguments (or at least recent citations in recognized journals), model-makers can vary ignition delay by over an order of magnitude.

As most of these published models have been ``validated'' and reproduce their target ignition delays quite accurately, there must be cancellation of errors occurring within each given model.
The parameters cannot be treated as independently verified and care must be taken when combining parameters from two or more models.
{This is especially true when some models omit some pathways altogether (implicitly assuming a rate of zero).}

It is tempting to use these substitutions to bring the model closer to the experimental data, for example by defining the fitness function such that the genetic algorithm minimizes the error metric, but this can take the model further from the truth; the analysis of reaction 293 shows one example.
Instead, we suggest a more suitable procedure is to use the tools demonstrated here to rank the  discrepancies by magnitude of impact, then have a ``blind'' reviewer (who does not know whether the substitution will worsen or improve the model fit) resolve the discrepancies through a literature search or, if needed, additional calculations.
Some models might appear to get worse though this process, but they will be more honest, and by revealing other masked errors this will be a step towards a more accurate set of detailed kinetic combustion models in the long run.

An interesting direction for future work would be to extend the analysis to different experimental targets, especially speciation data (showing intermediate species concentrations in either ignition or flow reactors), which are likely to constrain the parameters more than global ignition delay times.
Concerted efforts to assimilate all kinetic models, parameters, and experimental measurements, into shared databases with common interfaces will greatly help resolve discrepancies such as those shown in this article.


\section*{Acknowledgements}
This material is based upon work supported by the National Science Foundation under Grant Nos.\ 1403171, 1605568, and 1535065;
the Women and Minorities in Engineering program at Oregon State University; and
the Department of Chemical Engineering at Northeastern University.

\section*{Availability of material}
The scripts described in this article, the figures, and the data and plotting scripts necessary to reproduce them, are available openly under the CC-BY license~\cite{supplementary}.

\bibliography{refs}
\bibliographystyle{elsarticle-num}


\end{document}